\documentclass[a4paper]{article}
\usepackage{arxiv}
\usepackage[utf8]{inputenc}
\usepackage{enumerate}
\usepackage{amsmath}
\usepackage{graphicx}
\usepackage{multirow}
\usepackage{caption}
\usepackage{url}
\usepackage{hyperref}
\usepackage{float}
\usepackage[super]{nth}
\usepackage[labelfont=bf]{caption}
\usepackage[labelsep=period]{caption}

\usepackage{algorithm} 
\usepackage{algpseudocode}

\usepackage{tikz}
\usetikzlibrary{backgrounds,fit,decorations.pathreplacing, shapes}  
\usepackage{tkz-graph}

\usepackage[font=scriptsize,labelfont=scriptsize]{subfig}
\usepackage{subfig}

\usepackage[super]{nth}
\usepackage{adjustbox}
\usepackage{nameref}

\usepackage{xcolor}

\usepackage{listings}

\lstdefinestyle{mystyle}{
    basicstyle=\footnotesize,
    breakatwhitespace=false,         
    breaklines=true,                 
    captionpos=b,                    
    keepspaces=true,                 
    numbers=left,                    
    numbersep=5pt,                  
    showspaces=false,                
    showstringspaces=false,
    showtabs=false,                  
    tabsize=2
}
\tikzstyle{vertex}=[circle,draw=black, fill=white,sloped,minimum size=17pt,inner sep=5pt]

\usepackage{tikz}
\usetikzlibrary{decorations.pathreplacing,calc}

\newcommand{\figref}[1]%
{Figure \ref{#1}%
}

\algnewcommand{\LineComment}[1]{\State \(\triangleright\) #1}

\makeatletter
\def\@copyrightspace{\relax}
\makeatother

\title{Inventory Management - A Case Study with NetLogo}

\author{Rui Portocarrero Sarmento \\
PRODEI - Faculty of Engineering, University of Porto\\
Rua Dr. Roberto Frias, s/n\\
4200-465 Porto, Portugal\\
mail@ruisarmento.com
}

\begin{document}

\maketitle


\begin{abstract}

Multi-Agent Systems (MAS) have been applied to several areas or tasks ranging from energy networks controlling to robot soccer teams. MAS are the ideal solution when they provide decision support in situations where human decision and actions are not feasible to operate the system in control and in real-time. Thus, we present a case study that is related to dynamic simulation of an automatic inventory management system. We provide two types of agents, the clients, and the seller agents. Through a system of communication, the agents exchange messages to fulfill their inventory needs. The client agents trade products in quantities according to their needs and rely on seller agents if other clients in the retailer chain cannot provide the needed items.
Additionally, it is expected that the trading between a client and the sellers is done through a reverted type of auction. This case study MAS uses BDI and FIPA-ACL in its implementation resulting in a clear simulation of the system. We expect to provide a comparison between two distinct situations. One first situation where there are no transactions between retail stores and only external transactions with providers and other situation where both internal and external transactions are allowed.     

\end{abstract}



\section{Introduction}

We are aware of small size retail chains where the store's managers execute stocks management of several stores but in a manual fashion, not automatically. The stores also have to communicate with the providers when in need of inventory. The manager also does this manually and doesn't use any auction to acquire products.

One of the contributions of this work is to establish automatic inventory control for the manager. We wish to provide electronic communication of items needs between supermarket agents. Each supermarket agent communicates with other supermarkets to check if they have an excess of some particular item in stock. Thus, if another agent has excess stock, it can provide the item to the chain agent in need of stock, lowering the costs with external providers and also the costs of item storage.

Moreover, we want the retail chain stores to establish communication with external providers that might automatically provide the best price from the provider. For this, we wish to create a reverse auction with providers.

Some ad hoc multi-agent systems are often created by researchers and developers without using frameworks. Nonetheless, some frameworks have arisen recently, and these frameworks implement common standards. Some of these standards like, for example, the FIPA agent system platforms and communication languages, have an important role for the framework user. These frameworks save developers time and also aid in the standardization of MAS development. Relevant surveys were previously written and provide insight over several frameworks available \cite{nikolai2009tools,allan}.

Another contribution of this work is to provide a contextualization of a hypothetical NetLogo user. BDI and FIPA-ACL are standards that provide a more structured way to program MAS, and this is now possible with NetLogo. Thus, by providing additional knowledge of the NetLogo features the user is empowered to do a better and more professional work with NetLogo. Additionally, the provided inventory management case study is done by applying introduced concepts in NetLogo. Therefore, there is needed contextualization.

Introducing this work structure, we start by providing contextualization in \ref{rel}. Thus, we provide significant related work in this area. Then, we describe our case study in \ref{case}. In \ref{res}, we exhibit results obtained from the simulation of the MAS system built and described in \ref{case}. In \ref{dis} we provide a discussion of the results we gathered from the simulations. Finally, we conclude our work, and we also provide some insight into future developments of this work in \ref{conc}. 
  
\section{Related Work}\label{rel}
    
In this section, we provide several milestones and previous work on the subject of MAS for inventory management and also the NetLogo framework. Thus, we introduce the earlier work on this concept and also the relevant features of NetLogo that will provide a solution for the development of a simulation scenario. 

\subsection{Inventory Management}

In \cite{DBLP:journals/simulation/Signorile02}, Signorile emphasizes that ``the dynamics of the enterprise and the market make the inventory management difficult: materials do not arrive on time, production facilities fail, and employees are ill, causing deviations from the plans. In some cases, these events may be dealt with locally''. In particular, Signorile demonstrated that agents ``can be used to control an important part of the supply chain, the inventory of a retail store. In fact, the agent model outperforms the current EOQ (Economic order quantity) model" that the retail institution where the case study took place, is using (2002).

In \cite{Signorile2005}, Signorile expands their previous work and their decision support system to the inventory system (excluding supply chain activities) for a medium-sized department store in the United States (2005).

In \cite{Jiang20096520}, Jiang and Sheng approach this problem in a supply chain with non-stationary customer demand. They use MAS with reinforcement learning proved experimentally to be effective (2009).

In \cite{6285832}, Boudouda and Boufaida mention that ``the agent-based model proposed consists of a set of subsystems agent, each subsystem is an actor in the SC (Supply Chain); these subsystems are grouped into several levels. They seek to coordinate and synchronize different activities. Each subsystem is structured by three cognitive agents (Purchaser Agent, manager stock agent and delivery agent), these agents interact with each other to accomplish their tasks and can communicate, negotiate and collaborate with other subsystems through a communication interface. Two types of interaction between subsystems are defined: internal interaction for the actors belonging to the same level, the second type external can occur between subsystems of different levels..". Additionally, the authors stress that their model ``provides a collaboration and negotiation between different actors in the supply chain. It does not replace existing tools and strategies for the SC, but it can be used as a supplement which improves it" (2012).

\subsection{NetLogo}

NetLogo is a platform independent environment used for modelization of social and natural phenomena as multi-agent systems. It provides means to simulate with a great number of agents. The NetLogo platform was developed by Uri Wilensky in 1999 and is under continuous development at the Northwestern's Center for Connected Learning and Computer-Based Modeling (CCL).

The CCL is a research group headed by Prof. Uri Wilensky. The center includes staff and students at Northwestern working together and in association with researchers from around the world. The group includes educational researchers, curriculum developers, software engineers, and model builders. The center is funded by Northwestern, the USA National Science Foundation, and a few commercial sponsors. 

Using the NetLogo language, students and researchers have constructed large numbers of models of complex phenomena in the natural and social worlds. The Models Library that comes with NetLogo covers phenomena in biology \cite{primeiro,Stamatopoulou07formalmodelling}, chemistry \cite{galic2014modeling,Liu201392}, informatics \cite{Blum:2015:ABS:2812169.2812170}, economics \cite{biondo2013}, sports \cite{primeirosports}, sociology \cite{wallentin2015agent,pluchino2014,Balaraman:2014:ENE:2685617.2685656,Dickerson:2014:MSN:2667369.2667395}, medicine \cite{seal2011agent,fekir2011segmentation} and a variety of other domains.

Although the NetLogo platform offers support for reactive agent systems, modeling BDI agents and explicit symbolic message exchange is not supported. Sakellariou et al. \cite{conf/setn/SakellariouKS08,sakellariou2008teaching,sakellariou2009mas} have extended NetLogo with libraries to support both and used it for setting coursework in an Intelligent Agents unit.

According to Sakellariou et al. \cite{sakellariou2008teaching}, in the context of an Agent and Multi-Agent Systems course, the student's demands poses an interesting problem regarding practice procedures. Educators have reported a variety of environments and techniques they use to increase active learning. The authors report their experience using NetLogo as part of the practical coursework that students need to carry out within an Intelligent Agents course. Also, Sakellariou et al. describe two extra NetLogo libraries provided to students, one for BDI-like agents and one for ACL-like communication.

Sakellariou et al., describe the two additional NetLogo libraries with more detail than previously in \cite{conf/setn/SakellariouKS08}. One for BDI-like agents and one for ACL-like communication that, according to the authors, provide effortless development of goal-oriented agents, that communicate using FIPA-ACL messages. The authors include one simulation scenario that employs these libraries to provide an implementation in which agents cooperate under a Contract Net protocol.

\subsubsection{BDI}

The BDI library for NetLogo provides several NetLogo reports and procedures to implement belief–desire–intention software model. 

The library is built using the standard programming language provided by NetLogo. The only requirements by the authors of the library are that \cite{Sakellariou}:

\begin{enumerate}
    \item ALL complex agents MUST have two declared -own variables:"beliefs"  and "intentions". These are the variables to which beliefs and intentions are recorded. So, in the users model if there is a breed of agents which you wish to model as "BDI" agents, then the user should have a BREED-own [beliefs intentions] declaration (along with any other variables the user wishes to include in his model). The authors also stress that when the user creates the variables, they should initially set their values to empty a list ([ ]). 
    \item The users also must have ticks in his model or timeout facilities of the library will not function.
    \item The user's model and interface should include a “switch” in the interface, named “show-intentions”. This is necessary since the code of the receive message procedure checks in each cycle whether to output the messages or not.
\end{enumerate}

In this section, we introduce and explain the most prevalent and essential code to understand other tasks described in this paper.

A belief in the library is a list of two elements: the type and the content \cite{Sakellariou}. 

\begin{itemize}
    \item The belief type declares the type of the belief, i.e., indicates a “class” that the belief belongs to. Examples could include any string, e.g. “position”, “agent”, etc. Types facilitate belief management.
    \item The belief content is the content of a certain type of belief. It can be any NetLogo structure (integer, string, list, etc.). The authors emphasize that there might be many beliefs of the same type with different content. However, two beliefs of the same type and content cannot be added. 
\end{itemize}

The following code creates a new  belief. (does not stores it in  beliefs memory): 

\begin{lstlisting}[ firstline=23, lastline=25]
to-report create-belief [b-type content]
  report (list b-type content)
end 
\end{lstlisting}

To enable adding information to the beliefs structure, the following code is provided by the library:

\begin{lstlisting}[ firstline=38, lastline=41]
to add-belief [bel]
  if member? bel beliefs [stop]
  set beliefs fput bel beliefs 
end 
\end{lstlisting}

To be possible to remove a belief from the list of beliefs the authors provided the following code:

\begin{lstlisting}[ firstline=43, lastline=45]
to remove-belief [bel]
 set beliefs remove bel beliefs 
end 
\end{lstlisting}

To update the information of some particular belief, the following code was created by the authors:

\begin{lstlisting}[ firstline=77, lastline=80]
to update-belief [bel]
   remove-belief read-first-belief-of-type belief-type bel
   add-belief bel
end
\end{lstlisting}

Regarding intentions, several pieces of code were developed by the authors to provide an abstraction for a BDI-like implementation of MAS in NetLogo.

The following code provides the execution of an agent intention:

\begin{lstlisting}[ firstline=86, lastline=93]
to execute-intentions
;;;locals [myInt]  ;; first intentions
  if empty? intentions [stop]
  let myInt get-intention
  run intention-name myInt
  if runresult intention-done myInt [remove-intention myInt]
  if show-intentions [set label intentions] ;; Just for debugging. 
end 
\end{lstlisting}

As implemented by the authors, the intentions are stored in a STACK of intentions. The first argument is the intention name that should be some executable procedure the library user encodes in NetLogo. The second argument should be a NetLogo REPORTER that when evaluates to true, the intention is removed (either accomplished or dropped). As the authors designed the library, both values have to be STRINGS. Thus, the following code adds an intention in the intentions list:

\begin{lstlisting}[ firstline=135, lastline=137]
to add-intention [name done]
  set intentions fput (list name done) intentions
end 
\end{lstlisting}

Finally, for the removal of a specific intention from the intention stack, we have the following code:

\begin{lstlisting}[ firstline=122, lastline=124]
to remove-intention [bdi-lib##intention]
  set intentions remove-item (position bdi-lib##intention intentions) intentions
end 
\end{lstlisting}

\subsubsection{FIPA-ACL}

This library for NetLogo implements FIPA-ACL like messages in NetLogo \cite{SakellariouFIPA}. In this section, we introduce and explain the most common and important code to understand other tasks described in this paper.

The library authors have some requirements for the user of the library in NetLogo:

\begin{enumerate}
    \item All agents that can communicate MUST have a declared -own variable incoming-queue. This is the variable to which all messages are recorded. So, in the user's model, if there is a breed of agents which the user desires to communicate, then the user should have a BREED-own [incoming-queue] declaration (along with any other variables that the user wishes to include in his model. The authors stress that when the user creates the variables, he should set its values to an empty list ([ ]).
    \item The user's model should include a “switch” named “show-messages” in the interface. The authors explain this is necessary since the code of the received message checks in each cycle whether to output the messages or not.
\end{enumerate}

The code to enable sending of messages provides the definition of the sender and a receiver(s). In agents simulation, some issues might arise from communications like, for example, if the agent that should receive the message is "killed". One solution to this problem is nothing happens. Other solution could yield an error message. An alternative would be to create a safe send. The following code implements the author solution \cite{SakellariouFIPA}:

\begin{lstlisting}[ firstline=30, lastline=37]
to send [msg]
  let recipients get-receivers msg
  let recv 0
  foreach recipients [
   set recv turtle (read-from-string ?)
   if recv != nobody [without-interruption [ask recv [receive msg]]] ;; read-from-string is required to convert the string to number
  ]
end
\end{lstlisting}

Message reception deals with updating incoming-queue:

\begin{lstlisting}[ firstline=41, lastline=44]
to receive [msg]
   if show_messages [show msg]
   set incoming-queue lput msg incoming-queue
end 
\end{lstlisting}

The following NetLogo reporter returns the next message in the list and removes it from the queue:

\begin{lstlisting}[ firstline=47, lastline=52]
to-report get-message
  if empty? incoming-queue [report "no_message"]
  let nextmsg first incoming-queue
  remove-msg
  report nextmsg     
end 
\end{lstlisting}

The following code represents an explicit remove-msg procedure. This is needed since reporters in NetLogo cannot change a variable's value: 

\begin{lstlisting}[ firstline=62, lastline=64]
to remove-msg
  set incoming-queue but-first incoming-queue
end 
\end{lstlisting}

To enable broadcasting to all agents of breed of type t-breed, the following code was provided by the library authors:

\begin{lstlisting}[ firstline=66, lastline=75]
;; broadcasting to all agents of breed t-breed
to broadcast-to [t-breed msg agent]
  foreach [who] of t-breed [
    ifelse ? = agent [
      ;;;do nothing
      ][
       send add-receiver ? msg 
      ]        
  ]
end 
\end{lstlisting}

NOTE: We introduced the exclusion of the agent that broadcasts the message since, as we will see later in this document, in the case of a client broadcasting to other clients we were not interested that the client agent emitted messages to himself. 

To create Messages and add the sender, the following reporter code is essential:

\begin{lstlisting}[ firstline=79, lastline=81]
to-report create-message [performative]
 report (list performative (word "sender:" who) ) 
end 
\end{lstlisting}

Finally, to provide means to add fields to a message the following pieces of code were created by the authors:

\begin{lstlisting}[ firstline=114, lastline=144]
;;; ADDING FIELDS TO A MESSAGE
;; Adding a sender to a message.
to-report add-sender [sender msg]
  report add msg "sender:" sender
end

;; add a receiver
to-report add-receiver [receiver msg]
  report add msg "receiver:" receiver
end

;; adding multiple recipients
to-report add-multiple-receivers [receivers msg]
  foreach receivers
  [
    set msg add-receiver ? msg
  ]
  report msg
end

;; Adding content to a message 
to-report add-content [content msg]
  report add msg "content:" content
end

;; Primitive Add command 
to-report add [msg field value]
  ifelse field = "content:"
  [report lput value lput field msg]
  [report lput (word field value) msg]
end  
\end{lstlisting}

We should stress that the previous code also provides means to add a primitive associated with the message. This, simultaneously with the other pieces of code provided by the authors, enables a FIPA-ACL like structure for our messages between agents.

\section{Case Study}\label{case}


We carried several experiments with the work of Sakellariou et al. \cite{conf/setn/SakellariouKS08,sakellariou2008teaching,sakellariou2009mas}.
We established a simulation environment with the following characteristics \footnote{Available Code at \url{https://github.com/Sarmentor/InventoryManagement}.}:

\begin{enumerate}
   
    \item Client Agents (retailer group agents) in their fixed locations and with an item database with the following information for each item
    \begin{enumerate}
        \item current item quantity
        \item minimum item quantity
        \item maximum item quantity
        \item price paid for item in last transaction
    \end{enumerate}
     \item Seller Agents (providers) with an items database with the following information for each item
    \begin{enumerate}
        \item current item price
        \item minimum item price
        \item maximum item price
    \end{enumerate}
\end{enumerate}

With these characteristics, we had an idea of simulating several client and seller agents. Thus, we needed to start the simulation with both type of agents database previously configured. Therefore, we had to state the initial state for each agent. 

\subsection{Initial State}

The initial state might be different for each client agent since we randomly select from 3 different item database. Thus, we vary the initial state of the client agents regarding item quantity in stock and price paid for each item in the last transaction.

One example of the initial conditions of a client agent database is available in table \ref{table1}.

\begin{table}[h!]
\centering
\captionof{table}{Client Initial State Example}
\begin{tabular}{ |c|c|c|c|c| }
\hline
\parbox[c]{1.5cm} {\raggedright item\_id } & stock & min\_stock & max\_stock & buy\_price \\
\hline
pao & 120 & 25 & 120 & 0.12 \\
leite 1lt & 100 & 10 & 100 & 0.54\\
bolachas & 50 & 15 & 50 & 0.8 \\
cerveja & 250 & 12 & 250 & 0.35\\
fraldas & 15 & 5 & 15 & 1.7\\
peixe & 20 & 2 & 20 & 2.75\\
carne & 30 & 5 & 30 & 2.1\\
\hline
\end{tabular}
\label{table1}
\end{table}

Additionally, the seller agents might also vary with five different initial conditions. Therefore, we change values from seller to seller, for example, regarding item price and also minimum item price.

Another example of initial conditions, this time for seller agent database is available in the following table:

\begin{table}[h!]
\centering
\captionof{table}{Seller Initial State Example}
\begin{tabular}{ |c|c|c|c| }
\hline
\parbox[c]{1.5cm} {\raggedright item\_id} & price & min\_price\_unit & max\_price\_unit \\
\hline
pao & 0.12 & 0.1 & 0.15 \\
leite 1lt & 0.54 & 0.45 & 0.65\\
bolachas & 0.7 & 0.5 & 0.8\\
cerveja & 0.35 & 0.27 & 0.45\\
fraldas & 1.5 & 1.3 & 1.9\\
peixe & 2.5 & 2.2 & 3.2\\
carne & 2.25 & 1.95 & 2.73\\
\hline
\end{tabular}
\label{table2}
\end{table}

After selecting initial conditions settings for each agent, we can start the simulation. The agents follow predetermined behaviors we describe in the next subsections. 

\subsection{Client Agents}

Succinctly, the Client Agents have the following behaviors:

\begin{enumerate}
    \item Appear randomly at a location in the environment
    \item listen to messages
    \begin{enumerate}
        \item if an ``cfp" was received from another client
        \begin{enumerate}
            \item checks if there are enough items in stock to provide the other client. If the answer is yes sends a message ``propose" to the other client with the item quantity and price.
            \item if there are not enough items available then it sends a ``refuse" message to the other client.
        \end{enumerate}
        \item if an ``accept-proposal" message was received:
        \begin{enumerate}
            \item update stocks and reduce the quantity of the available item in the database
        \end{enumerate}
        \item if no message was received continues to 3.
    \end{enumerate}
    \item simulate selling items to virtual buyers and in random quantities
    \item check inventory
    \begin{enumerate}
        \item if an item stock reached its possible minimum and internal trading is not allowed, then jumps to 4(a)v. If it is allowed then:
        \begin{enumerate}
            \item broadcasts {\em cfps} message to other retail clients in the group
            \item collects proposals
            \item adds beliefs ``proposal-from-clients"
            \item evaluates beliefs (``proposal-from-clients")
            \begin{enumerate}
            \item if the proposal was accepted then:
                \begin{itemize}
                    \item sends ``accept-proposal" message to the client with best proposal.
                    \item updates stock by increasing the quantity of the item provided by the other client agent and returns back to 2.
                \end{itemize}
            \item if all clients sent ``refuse" message then jumps to 4(a)v.
            \end{enumerate}
            \item broadcasts {\em cfps} message to sellers
            \item collects proposals from sellers
            \item adds beliefs ``proposal"
            \item evaluates beliefs (``proposal")
            \begin{enumerate}
                \item if the proposal has the best price and was accepted after some specified quantity of auction rounds then:
                \begin{itemize}
                     \item sends ``accept-proposal" message to the winning seller.
                     \item updates item stock in database after receiving ``success" message from wining seller.
                \end{itemize}
                \item if the proposal was not accepted then back to 2.
            \end{enumerate}
        \end{enumerate}
        \item if no item stock reached its admissible minimum returns to 2.
    \end{enumerate}
\end{enumerate}

\subsection{Seller Agents}

Succinctly, the Seller Agents have the following behaviors:

\begin{enumerate}
    \item Listen to {\em cfps} from clients
    \item Evaluate and reply proposal with the price for the item. This price is a random number between the asked price and the seller's minimum admissible price for the item 
    \item Listen to replies from clients, if its a new cfp then 2 or else:
    \begin{enumerate}
        \item if the proposal was accepted (primitive ``accept-proposal") then:
        \begin{enumerate}
            \item Sends items to client
            \item Updates current price for item
        \end{enumerate}    
        \item if the proposal was not accepted (primitive ``reject-proposal") then back to 1.
    \end{enumerate}
\end{enumerate}

\subsection{Auction System}

When the client agents need to be provided by external agents, we will need to implement an auction system so the offered price for the items would be the best possible. As the external providers have different rates, it is mandatory to establish an auction when the ``cfp" messages are broadcasted. Thus, we choose to create a reverse auction, and the following workflow is executed:

\begin{enumerate}
    \item Client A needs an item and broadcasts ``cfp" messages to all seller agents with the current needed item and its price. Two different situations can happen:  
    \begin{enumerate}
        \item If all sellers are not able to provide the item with the requested price they refuse to propose and the auction is interrupted
        \item If sellers can provide the item they will propose their best price. This best price is a random value between the asked price in the current auction round and the minimum acceptable price for the provider/seller.
    \end{enumerate}
    \item Client A then broadcasts yet another set of ``cfp" to the sellers with the best price in 1(b).
    \item Client A then selects the best offer after a predefined amount of auction rounds or if there is only one seller biding. If there is a tie of price between sellers, Client A selects the first seller to have provided the best price.
\end{enumerate}

\section{Metrics for Evaluation}

There are two possible situations in our MAS system:

\begin{enumerate}
    \item Only external transactions between clients and sellers
    \item Both internal (between clients, i.e., the agents belonging to the retail chain) and external transactions (between retail chain agents, i.e., the clients, and the seller agents)
\end{enumerate}

For the evaluation of our inventory management MAS we decided that for the comparison of both situations we should measure several metrics:

\begin{itemize}
    \item Average item price, for the clients, given by the formula:
    $$AIP = \frac{\sum_{i=1}^{k} ItemTradePrice_{i}}{k}$$
    \item Average time a transaction takes, from consulting the internal/external agents until stock update if transaction is successful, given by the formula:
    $$AITT = \frac{\sum_{i=1}^{k} ItemTradeTicks_{i}}{k}$$
    \item Internal/external ratio (Only in case of internal and external buying), given by the formula:
    $$ITR = \frac{Total Internal Trades}{Total External Trades}$$
\end{itemize}

With these measures, we expect to have an idea of how these metrics evolve over the simulation period. These are generic measures that might provide some insight into the system behavior, nonetheless, as we will see, the initial conditions of the client agents influence the evolution of the simulations.

\section{Results}\label{res}

In this section, we will provide the results of our simulations. We choose to simulate with five external providers (sellers) and a retail chain of two agents (the clients).

\subsection{External Trading Only}

\begin{figure}[H]
  \centering
  \includegraphics[scale=0.2]{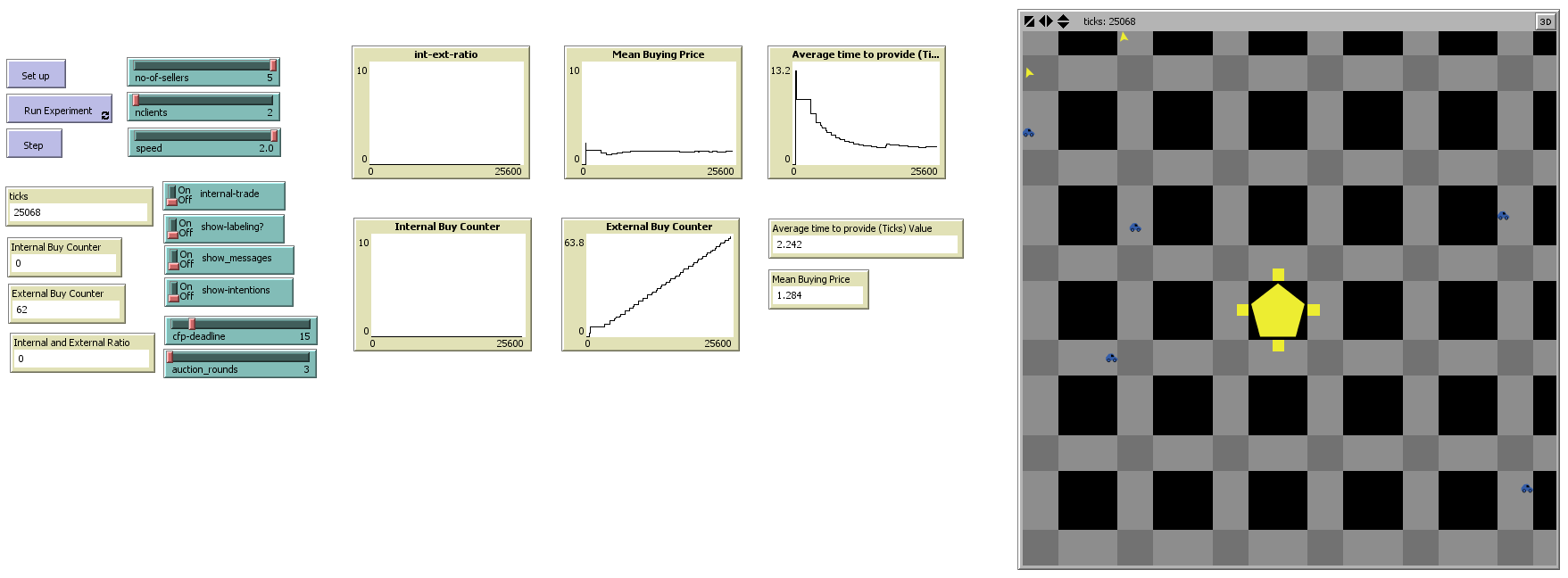}
  \caption{Interface with external trading mode simulation}
  \label{fig:interface}
\end{figure}

The external transactions only mode, for the clients, imply we have no internal transactions. \figref{fig:interface} captures the NetLogo interface of such simulation. 

\begin{figure}[H]
  \centering
  \includegraphics[scale=1]{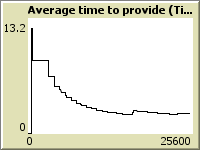}
  \caption{Interface with external trading mode simulation}
  \label{fig:time}
\end{figure}

In this mode, the results we obtained were stable after around 25000 ticks of simulation. This translated into 62 external buying transactions for the client agents. \figref{fig:time} provides evidence of the decreasing of ticks it takes for a client to be provided in this mode. It is clear that the implementation of the reverse auction gets faster as the simulation evolves.

This is explained by the fact that the clients register the best prices after the first auction is finished. Additionally, when they need to buy the same product later, they request a lower price to the seller agents which makes further auctions successful in fewer auction rounds, thus, the value of 2.242 ticks when the simulation stabilizes.

Regarding the average buying price, the value stabilizes at around 1.284. This value starts higher at the beginning of the simulation but, similarly to the previous results of this mode and for the same reasons, it lowers as the simulation evolves.

\subsection{Internal and External Trading}

Depending on the initial conditions, the simulation has different results and takes a different number of ticks to get to a stable state. For example, with an initial state with one client agent with the maximum number of items available for all the products, and the other agent with low amounts of items, we can expect to have more internal trades than external trades for a long period or many ticks.

Thus, in these conditions and for approximately 3000 ticks, the client agents did 1238 internal buys and 0 external buys. Additionally, the average number of ticks it takes for a client to be provided is 0 which is the main advantage of having only internal trades. This is explained by our number of client agents which is low and the fact that we do not execute any auction in this situation of only having internal trading.

Regarding the average buying price, the value is 1.373. Compared with another mode where we can only have external trades, this value is higher which is expected as we do not have external transactions that rely on the reverse auction. This leads clients to be provided with the lowest price possible, thus lowering the average buying price with external trading.

\begin{figure}
  \centering
  \includegraphics[scale=1]{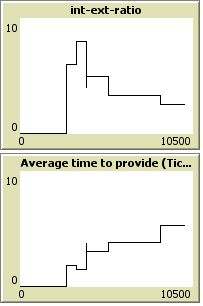}
  \caption{Time to Provide and Internal/External Ratio with internal and external trading mode simulation}
  \label{fig:plots}
\end{figure}

We did additional simulations, this time with other initial conditions. With unbalanced client's stocks, for example, with maximum stock for some items and minimum for others, there is more chance for a balance between internal and external trading. \figref{fig:plots} shows that, with 10000 ticks of simulation, ten internal buys, and four external solicitations, an interesting case is simulated. 

It is visible that as the internal/external ratio decreases the time it takes until the client providing finishes are higher. This means that initially in the simulation, more internal transactions were done than external transactions. Then, further in the simulation, some external transactions occur which increases the time needed to provide the clients with the items they need. At this point, the average number of ticks it takes for a client to be provided is 5.286. Additionally, regarding the average buying price, the value is 0.7. This value is lower than previous simulations; this might be due to the items that were sparse in the initial conditions and their lower price. Thus, with more simulation ticks, this value is expected to increase as more expensive items are bought.

\begin{figure}[H]
  \centering
  \includegraphics[scale=1]{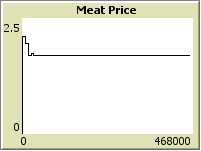}
  \caption{Meat item price evolution}
  \label{fig:meat}
\end{figure}

\begin{figure}
  \centering
  \includegraphics[scale=1]{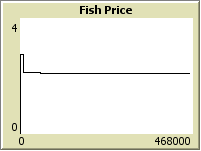}
  \caption{Fish item price evolution}
  \label{fig:fish}
\end{figure}

Additionally, as the simulation progresses, there are more auctions, the price of the items naturally lowers. \figref{fig:meat} and \figref{fig:fish} shows this situation. 

Both meat and fish price in this simulation starts higher. This is due to the internal transactions at higher initial prices for both items. As the simulation evolves more and more external transactions occur. Thus, with more auctions happening, there is a natural progression of prices towards the minimum possible price regarding the providers in the simulation.

An example of a provider context after several thousands of simulation ticks is provided in the following table 3.

\begin{table}[h!]
\centering
\captionof{table}{Seller Final State Example}
\begin{tabular}{ |c|c|c|c| }
\hline
\parbox[c]{1.5cm} {\raggedright item\_id} & price & min\_price\_unit & max\_price\_unit \\
\hline
pao & 0.15 & 0.13 & 0.14 \\
leite 1lt & 0.5 & 0.5 & 0.54\\
bolachas & 0.6 & 0.6 & 0.8\\
cerveja & 0.3 & 0.29 & 0.37\\
fraldas & 1.6 & 1.45 & 1.75\\
peixe & 3.1 & 2.9 & 3.3\\
carne & 1.7 & 1.7 & 2.8\\
\hline
\end{tabular}
\label{table2}
\end{table}

In this table, taken from a simulation after several external transactions, it is clear that for example the item "leite 1lt" as well as "bolachas" and "carne" reached the lowest price possible in the simulation. Recording the initial prices for this provider, we had, at the beginning of the simulation, the following table 4.

\begin{table}[h!]
\centering
\captionof{table}{Seller Initial State Example}
\begin{tabular}{ |c|c|c|c| }
\hline
\parbox[c]{1.5cm} {\raggedright item\_id} & price & min\_price\_unit & max\_price\_unit \\
\hline
pao & 0.15 & 0.13 & 0.14\\
leite 1lt & 0.5 & 0.5 & 0.54\\
bolachas & 0.6 & 0.6 & 0.8\\
cerveja & 0.3 & 0.29 & 0.37\\
fraldas & 1.6 & 1.45 & 1.75\\
peixe & 3.1 & 2.9 & 3.3\\
carne & 1.8 & 1.7 & 2.8\\
\hline
\end{tabular}
\label{table2}
\end{table}

Both these tables make us conclude that this provider had already the lowest prices possible for "leite 1lt" as well as "bolachas". Nonetheless, the item meat, designated by "carne" reached its lowest price of 1.7 after the simulation.

\section{Discussion}\label{dis}

With these results, it is clear that initial client states regarding item stocks influence, on a large scale, the evolution of the simulation. Therefore, it is expected that due to the relative randomness of initial states the simulation provides less similar results. Nevertheless, the implemented system provides a testbed for further improvements, and the preliminary results are logical and seem intuitive.

In a real situation, other factors should be regarded, like for example, the costs with item storage, the costs of transportation of items between clients and also the validity of products. 

Additionally, the providers (seller agents) should also have stock limitations as in real situations. Thus, the reduction of stock and, therefore, the capacity to provide or not a client agent should be studied. These factors would improve further the realism of the simulations and could provide better decision support to the manager of the retail chain group.

\section{Conclusions and Future Work}\label{conc}

With this work, our goal was to provide an introduction to the inventory management problem and how to approach it with MAS frameworks, more specifically, with the NetLogo and its expanded libraries. These libraries help to implement BDI and FIPA-ACL based MAS. We provided the most important concepts of these libraries.

Thus, by using these tools, we developed a trading system where we could focus on the inventory management performed by agents that could communicate with each other.

We provided results by exposing the results of this communication. We used several metrics to compare two situations of trading, with and without internal trading between agents belonging to the retail chain. Furthermore, we implemented a reverse auction when the agents belonging to the retail chain would need to buy from external providers.

We conclude that, by using these libraries, we increase comprehensibility and efficiency in the development and implementation of this MAS system. 

Additionally, by exploring the MAS system, we can easily simulate a complex system. The expected complexity of studying a real and similar system makes it imperative to use models that provide a test bench for different situations. These conditions would be very hard to test in real systems, and this is one of the main advantages of MAS.

\section*{Acknowledgments}
This work was fully financed by the Faculty of Engineering of the Porto University. Rui Portocarrero Sarmento also gratefully acknowledges funding from FCT (Portuguese Foundation for Science and Technology) through a Ph.D. grant (SFRH/BD/119108/2016).

\bibliographystyle{plain}
\bibliography{TNE}

\newpage






\end{document}